\documentclass[twocolumn]{emulateapj}

\usepackage{apjfonts}

\newcommand{\etal}{{\it et al.} }

\newcommand{\asca}{{\it ASCA} }

\newcommand{\xmm}{{\it XMM-Newton} }

\newcommand{\chandra}{{\it Chandra} }

\newcommand{\rxte}{{\it RXTE} }

\newcommand{\hetg}{{\it HETGS} }

\newcommand{\fekalfa}{{Fe~K$\alpha$} }

\newcommand{\bsax}{{\it BeppoSAX} }

\newcommand{\ic}{IC~4329A }

\newcommand{\fekb}{Fe~K$\beta$ }

\newcommand{\felya}{Fe~{\sc xxvi} Ly$\alpha$ }
\newcommand{\felyap}{Fe~{\sc xxvi} Ly$\alpha$}

\newcommand{\felyb}{Fe~{\sc xxvi} Ly$\beta$ }

\newcommand{\fexxvf}{Fe~{\sc xxv} (f) }
\newcommand{\fexxv}{Fe~{\sc xxv} }
\newcommand{\fexxvp}{Fe~{\sc xxv}}

\begin{document}

\title{HIGH RESOLUTION X-RAY SPECTROSCOPY OF THE Fe K COMPLEX IN IC~4329A}

\author{Barry McKernan\altaffilmark{1,2} and Tahir Yaqoob\altaffilmark{2,3}}

\altaffiltext{1}{Present Address: Department of Astronomy, 
University of Maryland, College Park, MD 20742}

\altaffiltext{2}{Department of Physics and Astronomy, 
Johns Hopkins University, Baltimore, MD 21218}
\altaffiltext{3}{Laboratory for High Energy Astrophysics, 
NASA/Goddard Space Flight Center, Greenbelt, MD 20771}

\begin{abstract}
We report the detection of complex Fe~K line emission 
from a \chandra High Energy Transmission Grating Spectrometer 
(\hetg) observation of the Seyfert~1 galaxy IC~4329A. The line emission
is double-peaked,
one peak centered at $\sim 6.3$~keV, and the other
at $\sim 6.9$~keV in the source rest frame. 
When modeled by Gaussians, the lower energy peak is resolved
by the HEG at $>99\%$ confidence, whilst the higher energy
peak is resolved at only $<90\%$ confidence. The best-fitting widths
are $\sim 21,000 \rm \ km \ s^{-1}$ and $\sim 4000 \rm \ km \ s^{-1}$
FWHM for the $\sim 6.3$~keV and $\sim 6.9$~keV peaks respectively.
If the peaks correspond to two distinct emission lines, then
the peak energies are redshifted with respect to the
expected line energies of Fe~{\sc i} K$\alpha$ and Fe~{\sc xxvi} Ly$\alpha$
by at least $650 \rm \  km s^{-1}$ and $950\rm \ km \ s^{-1}$ respectively.
Alternatively, the Fe~K line profile may be due to a single line
from a relativistic accretion disk. In that case the inclination angle
of the disk is required to be $24^{+9}_{-1}$ degrees, the outer radius
constrained to several tens of gravitational radii, and the radial
line emissivity flatter than $r^{-0.7}$.
Another possibility is that both peaks are due to distinct lines but each one relativistically
broadened by a disk. In that case the lower energy peak could
correspond to emission from Fe in a low ionization state, and the
high-energy peak to \felya emission. Then, the inclination angle is even less,
restricted to a few degrees. However, the radial emissivity law is allowed
to be steeper ($\sim r^{-2.5}$) and the outer radius does not have to be
fine-tuned. Yet another scenario is that the lower energy peak originates
in a disk but the higher energy peak originates in more distant matter.
The disk inclination angle is then intermediate between the last two cases
but the emissivity is again required to be flat.
We cannot rule out \fexxv He-like absorption modifying the observed line
profile. However, the data, and inferred emission-line parameters,
are insensitive to the presence of a Compton reflection continuum.
Including Compton reflection does, however, allow a steeper 
radial emissivity law for the relativistic line.
Future missions such as {\it Astro-E2} will be able to break
a lot of the degeneracy in the physically distinct models that
can all account for the \chandra data. Since IC~4329A is one of the
brightest Seyfert~1 galaxies it should be a good astrophysical laboratory
for studying the ionization structure of accretion disks around supermassive
black holes.
\end{abstract}

\keywords{galaxies: active -- galaxies: individual (\ic) -- galaxies: 
Seyfert -- techniques: spectroscopic -- X-rays: line -- emission: 
accretion -- disks: galaxies}

\section{Introduction}
\label{sec:intro}

Some of the Fe~K$\alpha$ fluorescent emission line in active galactic nuclei 
(AGNs) is believed to originate in a relativistic accretion disk around 
a supermassive black hole (see e.g. reviews by Fabian \etal 2000; 
Reynolds \& Nowak 2003).  Early studies of type~1 AGN using X-ray 
observations with \asca revealed a wide variety of Fe~K$\alpha$ line 
widths and shapes (e.g. Nandra \etal 1997;
Lubi\'{n}ski, P. \& Zdziarski 2001; Perola \etal 2002;
Yaqoob \etal 2002).  These results have been 
confirmed with the advent of the \chandra and \xmm X-ray 
observatories, but a more complex picture is emerging. New, high 
resolution observations are revealing that a narrow Fe~K line 
($<10,000$ km $\rm{s}^{-1}$ FWHM) component is common in type~1 AGN 
(Yaqoob \etal 2001; Kaspi \etal 2002; Pounds \etal 2001; Turner \etal 
2002; Fang \etal 2002; Reeves \etal 2002; Yaqoob \& Padmanabhan 2004), 
often superimposed on top 
of a broad ($\sim$ 10,000--100,000 km $\rm{s}^{-1}$ FHWM) component. 
Traditionally, narrow Fe~K lines are associated with matter at least 
several thousand gravitational radii from the black 
hole (one gravitational radius, $r_{g}$, defined as
$GM/c^{2}$). However, there is increasing evidence to suggest that narrow 
Fe~K lines can originate in, or have a significant contribution from, 
the putative accretion disk (Petrucci \etal 2002; Lee \etal 2002; Yaqoob \etal 
2003a). Narrow Fe~K lines originating from a disk can be interpreted 
in terms of truncated disk emission (e.g. Done, Madejski, \& \.{Z}ycki 2000) or 
in terms of a flattened radial line emissivity where the intensity 
per unit area falls off less steeply than $\sim r^{-2}$ (Yaqoob \etal 
2003a). Alternatively, the narrow Fe~K lines could be due to localized 
emission from the disk (Turner \etal 2002; Yaqoob \etal 2003a).
There are, of course, other possible interpretations of
the origin of narrow Fe~K lines (e.g. Sulentic \etal 1998; Elvis 2000; 
Hartnoll \& Blackman 2000).

Earlier observations with \asca also identified Fe K-shell emission due 
to highly ionized (H-like) Fe in two type~1 AGN, PG 1116+215 (Nandra \etal 
1996) and Ton S 180 (Turner \etal 1998; Comastri \etal 1998). Observations 
with the \chandra 
high energy transmission grating spectrometer 
({\it HETGS}) and the \xmm CCD detectors are now revealing Fe~K emission 
due to H-like and He-like Fe in other AGN 
(e.g. Reeves \etal 2002; Dewangan 2002; Page, Davis, \& Salvi 2003; 
Yaqoob \etal 2003a). As more examples of 
highly ionized Fe spectral features are detected in the spectra of 
type~1 AGN, it is clear that we need a better understanding of the 
complex ionization physics of the accretion disk (e.g. Nayakshin 
\& Kallman 2001; Ballantyne, Ross \& Fabian 2001; Ballantyne \& Ross 2002). 

Here we report on the spectral results of a \chandra \hetg 
observation of the type~1 AGN IC~4329A, in which we observe
a complex, broad, double-peaked
line profile. Preliminary results of this observation were 
presented in Yaqoob \& Padmanabhan (2004) and we now
present detailed modeling of the
Fe~K line profile at the highest spectral resolution yet available
(FWHM $\sim 1860 \ \rm km \ s^{-1}$ at 6.4~keV).
\ic is a bright Seyfert~1.2 
AGN at $z=0.016054$ (Willmer \etal 1991) embedded in a nearly edge-on 
host galaxy. It is one of the brightest Seyfert galaxies known, and
is sometimes brighter than 3C~273 in the 2--10 keV band.
\ic is luminous, with $L$~(2--10keV) $\sim 4 \times 10^{43}$ 
ergs $\rm{s}^{-1}$ typically, where we have converted from the luminosity 
given by Reynolds (1997) using $H_{0}=71$ km $\rm{s}^{-1} {\rm Mpc}^{-1}$ 
(Spergel \etal 2003) and $q_{0}=0$. 

\ic was observed previously with \asca (Mushotzky \etal 1995; 
Cappi \etal 1996), {\it GRO} (Madejski \etal 1995), \bsax 
(Perola \etal 1999), \asca simultaneously with \rxte (Done, 
Madejski \& \.{Z}ycki 2000) and \xmm simultaneously with \bsax 
(Gondoin \etal 2001). A Gaussian model fitted to the \asca observations 
(Done, Madejski \& \.{Z}ycki 2000) revealed strong (EW$=180\pm 50$ keV), 
broad (FWHM=$42,960 \pm 11,015$ km $\rm{s}^{-1}$) Fe~K emission 
peaking near $\sim 6.4$~keV.
The \xmm observations (Gondoin \etal 2001) revealed a narrow Fe~K$\alpha$ 
line (FWHM $<8785$ km $\rm{s}^{-1}$, EW$=43 \pm 1$ eV) 
originating in matter with a low ionization state
(the line peaked at $E=6.42^{+0.04}_{-0.03}$ keV). 

The paper is organized as follows.
In \S\ref{sec:obs} we discuss the 
observation and data. In \S\ref{sec:modeling} we discuss 
fitting different physical models 
of the complex spectrum in the Fe~K band. In 
\S\ref{sec:discussion} we discuss the implications of our results
and present our conclusions. 

\begin{figure}[h]
\epsscale{0.8}
\plotone{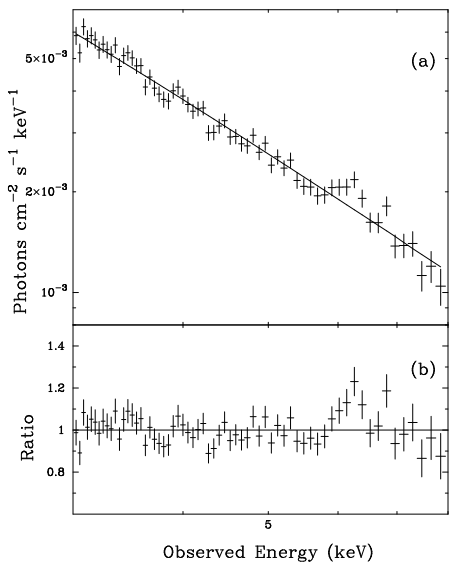}
\caption{\ic HEG spectrum between 3.0 and 8.0 keV binned at 0.16\AA\ 
compared with the best-fitting power-law model. (a) Photon spectrum, 
(b) Ratio of the data to the model.} 
\end{figure}

\section{Observation and Data}
\label{sec:obs}
\ic was observed with the \chandra \hetg (Markert \etal 1995)
on 2001 August 26 from UT 06:14:07 for 
a duration of $\sim 60$~ks.
The \chandra \hetg consists of two grating assemblies, a high-energy 
grating (HEG) and a medium-energy grating (MEG). We used only the combined 
$\pm1$ orders of the grating data. The mean \chandra total HEG and MEG 
count rates were 1.158 $\pm$ 0.0051 ct/s and 2.244 $\pm$ 0.0067 ct/s
 respectively. 
The source flux showed little variability over
the entire duration of the campaign. For example,
for HEG plus MEG combined first-order lightcurves binned at 128 s,
the excess variance above the
expectation for Poisson noise (e.g. see Turner \etal 1999)  was 
$(3.4 \pm 18.2) \times 10^{-5}$,
consistent with zero. Only single spectra per instrument were 
therefore extracted.

We made effective area files (ARFs or \emph{ancillary 
response files}), photon spectra and counts spectra following the 
method of Yaqoob \etal (2003b). 
A net exposure time of 59,087 s was obtained
(including a deadtime factor of 0.0162).
The HEG bandpass is $\sim 0.8$--10~keV and the 
MEG bandpass is $\sim 0.5$--10~keV but the effective area of both instruments 
falls off rapidly at either end of the bandpass. Background was not 
subtracted since it is negligible in the energy region of interest. 
The HEG has twice the spectral resolution (0.012\AA\ FWHM) of the MEG 
(0.023\AA\ FWHM) and has a larger effective area than the MEG in the
 Fe~K band so our analysis will focus on the HEG data,
since we are interested in studying the Fe~K region with
the highest spectral resolution currently available. The MEG spectrum 
of \ic is discussed elsewhere (McKernan, Yaqoob, \& Reynolds 2004, in 
preparation). 

\section{Modeling of the Fe~K Complex in \ic}
\label{sec:modeling}
We used XSPEC v.11.2.0 for spectral fitting to the HEG spectrum in 
the 3--8~keV band. This choice for the lower end of the bandpass avoids
the known complexities in the soft X-ray spectrum due to 
absorption and the higher end ensures that background
will indeed be negligble since the instrument efficiency area vanishes
rapidly at high energies. Galactic absorption of 
$4.55 \times 10^{20} \rm{cm}^{-2}$ 
(Elvis \etal 1989) has a negligible effect in this energy range and so 
was not included in our modeling. There has been a well-documented 
degradation in the quantum efficiency of \chandra ACIS due to molecular 
contamination.  However, the effect of this degradation is most pronounced 
at energies below $\sim 1$ keV and is negligible over the energy range discussed here
so we do not include ACIS degradation in our modeling. The 
$C$--statistic (Cash 1976) was used for minimization in all models and 
all statistical errors quoted are 90$\%$ confidence for one 
interesting parameter ($\Delta$C=2.706), unless otherwise stated. 

First we investigated how the 3.0--8.0 keV HEG data compared with a simple 
model consisting of a single power law plus intrinsic 
absorption, with the spectrum binned at 0.16\AA. We included
absorption in this initial fit, just to ensure that it can
be neglected in the bandpass that we used. 
The best--fitting power-law index for this model was 
$1.66^{+0.09}_{-0.05}$ and the absorbing column density was zero
with an upper limit of $5.1 \times 10^{21} \rm \ cm^{-2}$.
Figure~1(a) shows the HEG photon spectrum of 
\ic with the model superimposed. Figure~1(b) shows the ratio of the data 
to the model.  The single power law is clearly a good overall fit to the 
hard X-ray spectrum but there is a broad double-peaked
feature in the Fe~K 
band. The peaks occur at $\sim 6.2$ keV and $\sim 6.8$ keV in the 
observed  frame, and the lower energy feature appears to be 
broader than the high-energy one. To investigate the Fe-K emission 
complex in more detail,
we used spectra binned at 0.005\AA, which corresponds approximately
to the $1\sigma$ HEG resolution. Since no absorption is required
in the 3--8 keV band, we used only a power law for the continuum.
Several models of the complex Fe K emission are discussed below.

\subsection{Double Gaussian Model}
\label{doublegauss}

We added two Gaussian emission-line components to the 
power-law continuum model in order to fit the $\sim$6.2 keV 
and $\sim$ 6.8 keV emission peaks.
We did not include a Compton reflection continuum in the fit
since we will show {\it post-facto} that the emission-line
parameters are insensitive to the presence or absence of
a Compton reflection continuum (see \S\ref{complexcont}).
The best-fitting parameters for 
this model are given in Table~1.  Note that in Table~1 we show the 
velocity shift from systemic when the lower and higher-energy peaks
are identified with Fe{\sc~i}~K$\alpha$ emission 
(at 6.400 keV in the rest frame) and H-like \felya emission (at 6.966 keV 
rest frame) respectively. Even when we take into account 
the systematic uncertainty of the HEG wavelength scale ($\sim \pm$420, 
$\pm$460 km $\rm{s}^{-1}$ at $\sim 6.2$ keV and $\sim 6.8$ keV 
respectively)\footnote{http://space.mit.edu/CXC/calib/hetgcal.html},
the low-energy feature (if Fe{\sc~i}~K$\alpha$ emission) 
is redshifted by $>650$ km $\rm{s}^{-1}$ and the higher energy feature 
(if \felya) is redshifted by $>950$ km $\rm{s}^{-1}$ (both are $90\%$ 
confidence lower limits). If the higher energy feature is due to \fexxv 
forbidden or \fexxv resonance emission, 
then the blueshift is $\sim 12,200$ km $\rm{s}^{-1}$ 
or $\sim 9,220$ km $\rm{s}^{-1}$ respectively, relative to systemic 
velocity. 

\begin{figure}[h]
\epsscale{0.8}
\plotone{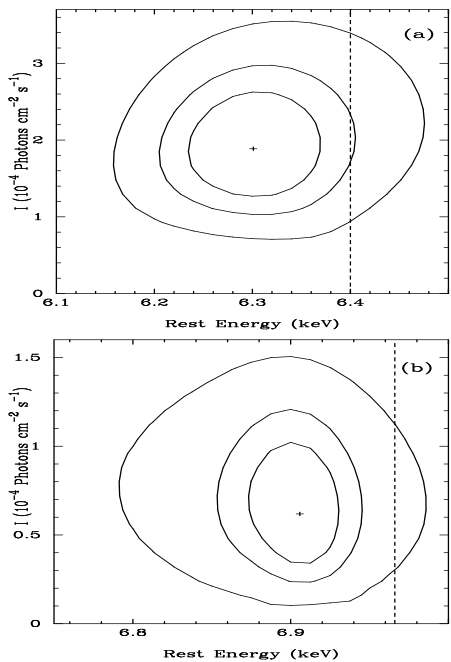}
\caption{Confidence contours of line intensity versus 
rest-frame energy for (a) the lower  energy Fe~K peak and (b) 
the higher energy 
peak from a double Gaussian model fit to the HEG spectrum of IC~4329A
(see \S \ref{doublegauss} and Table~1). 
The two-parameter joint confidence levels correspond to 68$\%$, 90$\%$, 
and 99$\%$ confidence. The dashed line in (a) 
denotes the rest-frame energy for Fe{\sc~i}~K$\alpha$ emission 
(6.400 keV) and that in (b) corresponds to the rest-energy of \felya 
emission (6.966 keV). 
The contours in (a) indicate that the lower energy feature 
is redshifted from the Fe{\sc~i}~K$\alpha$ rest-energy at $\sim 90\%$ 
confidence. From (b) it appears that the higher energy feature is 
redshifted from the rest-energy of the \felya emission line.}
\end{figure}

The low-energy Fe~K feature is moderately broad 
(FWHM=$20,825^{+10,110}_{-7375} \ \rm km \ s^{-1}$), as we saw from 
Fig.~1, whereas the higher energy feature is narrower 
(FWHM=$3680^{+5200}_{-2455}$ km $\rm{s}^{-1}$).  We estimated the 
significance of each feature by 
comparing the values of the $C$--statistic obtained
from fitting a power-law continuum only and from adding
the feature in question to the power-law only fit. The
resulting values of $\Delta C$ are shown in Table~1.
The addition of each Gaussian component adds three
free parameters to a fit, so 
the lower energy Fe~K feature was detected with a confidence 
level $>$ 4 $\sigma$ and the higher energy feature was detected at 
only $> 95\%$ confidence.  

\begin{figure}[h]
\epsscale{0.6}
\plotone{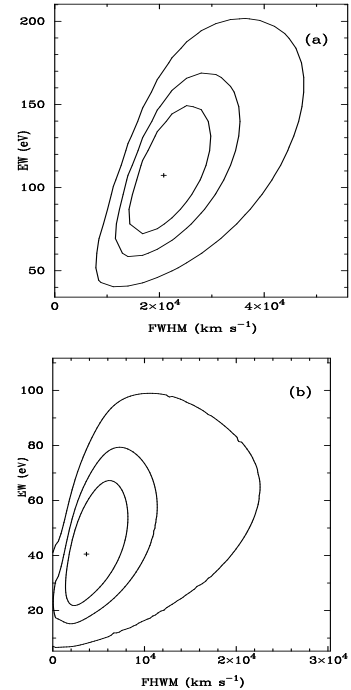}
\caption{Confidence contours of equivalent width (EW)
versus FWHM for (a) the lower energy 
Fe~K emission peak and (b) the higher energy peak from a double-Gaussian 
model fit to the HEG spectrum of IC~4329A
(see \S \ref{doublegauss} and Table~1). The two-parameter 
joint confidence levels correspond to 68$\%$ (innermost), 90$\%$ and 
99$\%$ (outermost).} 
\end{figure}

Fig.~2(a) shows a two-parameter joint 
confidence contour of line intensity versus rest-frame energy for the lower
energy 
Fe~K emission and  Fig.~2(b) shows a similar plot for the higher energy 
emission. Consistent with the spectral fitting results,
it can be seen that the lower energy Fe~K emission peak is marginally redshifted 
from the rest-frame energy of Fe{\sc~i}~K$\alpha$
(dashed line) at $\sim 90\%$ confidence. If we identify the 
higher energy feature with \felya emission, this line is also 
marginally (at $> 90\%$ confidence) redshifted with respect to systemic velocity 
(dashed line).  Fig.~3(a) shows a two-parameter joint confidence 
contour of the equivalent width (EW) versus FWHM for the lower energy
Fe~K emission and 
Fig.~3(b) shows a similar plot for the higher energy emission peak. 

\begin{figure}[h]
\epsscale{0.58}
\plotone{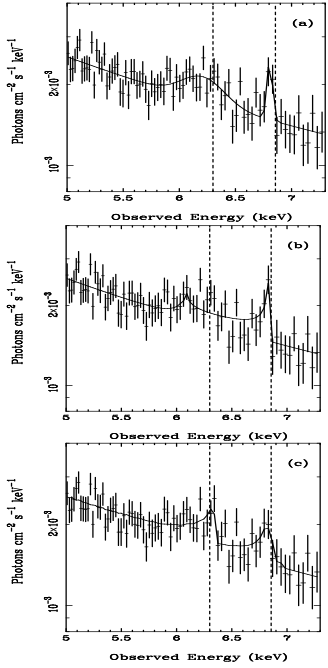}
\caption{\ic HEG photon spectra between 5.0 and 7.3 keV 
(observed energy), binned at 0.01\AA. The solid lines correspond to 
physically different
(and degenerate) models (see text for details). Note that the  
data are {\it not} unfolded: the spectra were made by multiplying
the ratio of the counts in each spectral bin to the predicted
model counts in that bin, by the best-fitting model. 
In each case the continuum is a simple power law and the emission-line
models are
(a) a double Gaussian (see \S \ref{doublegauss} and Table~1);
(b) a single disk line (see \S \ref{singlediskline} and Table~2), 
and (c) a double disk line (see \S \ref{doubledisk} and Table~3).
Dashed lines indicate the observed-frame energies expected for 
Fe{\sc~i}~K$\alpha$ (6.400 keV) and \felya (6.966 keV) emission 
respectively.} 
\end{figure}

Fig.~4(a) shows a close-up of the best-fitting Gaussian model fit 
(see Table~1) to the \ic HEG spectrum.  Again,
the offset of the Gaussian peak energies 
from the dashed lines in Fig.~4(a), which mark the energies of 
 Fe{\sc~i}~K$\alpha$ and \felya emission in the observed frame 
(6.299keV and 6.856keV respectively), show that both the lower and higher
energy peaks (if Fe{\sc~i}~K$\alpha$ and Fe~{\sc xxvi} Ly$\alpha$ 
respectively) are indeed marginally redshifted in IC~4329A. If the 
higher energy feature is \felya emission, then under some circumstances 
we should also expect significant Fe~{\sc xxvi} Ly$\beta$ emission (e.g. see 
Bautista \etal 1998; Bautista \& Titarchuk 1999). We therefore added 
a third Gaussian component to the model in Table~1 and we
extended 
the HEG data energy range to 3.0--8.3~keV since \felyb occurs at 8.251 keV 
in the rest-frame (for all successive models we shall return to fitting 
the data range 3.0--8.0 keV). \felyb is hardly detected 
($\Delta C$ improves by only $\sim 1$ for the addition of a Gaussian with 
three free parameters, and EW$<49$eV). The value of the 
Fe~Ly$\beta$:Ly$\alpha$ ratio is a powerful diagnostic since it 
can determine the process of excitation.  However, the measurement 
errors on the Fe~Ly$\beta$:Ly$\alpha$ ratio in these data are too large to 
obtain any meaningful constraints. On the other hand, since
\ic is one of the brightest Seyfert galaxies (sometimes it is brighter
than 3C~273), it is one of the best candidates to constrain
the Fe~Ly$\beta$:Ly$\alpha$ ratio with future missions such as {\it Astro-E2}.
We obtained the observed-frame 2--10~keV flux (for comparison with 
values in the literature)
for the \chandra HEG spectrum by extrapolating the above best-fitting
model down to 2~keV and up to 10~keV, and this yielded a value of
$16.2 \times 10^{-11} \ \rm erg \ cm^{-2} \ s^{-1}$, corresponding
to an intrinsic, source-frame 2--10~keV luminosity of $9 \times 10^{43}
\ \rm erg \ s^{-1}$.

Since we detected an emission line peak that
could be 
 due to \felya we searched 
for evidence of other features due to highly ionized Fe (as detected 
e.g. in NGC~7314 by Yaqoob \etal 2003a) by adding a third Gaussian 
model component. We find EW$<17$eV for \fexxvf (at the same offset velocity 
and the width of the higher energy Fe~K feature in Table~1 ) and we find 
marginally significant, narrow \fexxvf emission at the
galaxy systemic velocity 
(detected at $>95\%$ confidence, since $\Delta$C improves by $\sim 5.6$ 
for the addition of one free parameter).

\begin{deluxetable}{lrr}
\tablecaption{Double Gaussian Model}
\tablecolumns{3}
\tablewidth{0pt}
\tablehead{
\colhead{ } & \colhead{Low-Ionization} & \colhead{High-Ionization}\nl
\colhead{} &\colhead{Fe K Peak} &\colhead{Fe K Peak} \nl}
\startdata
E(keV)&$6.301^{+0.076}_{-0.073}$ & $6.906^{+0.028}_{-0.037}$\nl
c$\Delta$E/$E_{0}$(km $\rm{s}^{-1}$)$^{a}$ &$4640^{+3420}_{-3560}$ &
$2585^{+1590}_{-1210}$\nl 
$\sigma$(eV) & $186^{+90}_{-66}$&
$36^{+51}_{-24}$\nl 
FWHM(km $\rm{s}^{-1}$) $^{a}$ &
$20,825^{+10,110}_{-7375}$ & $3680^{+5200}_{-2455}$\nl 
$I \ ^{b}$ & $19.0^{+8.0}_{-6.9}$&
$6.2^{+4.4}_{-3.1}$ \nl 
EW(eV)&$110^{+46}_{-40}$ & $42^{+30}_{-21}$\nl
$\Gamma$&$1.76 \pm 0.06$ & $\ldots$ \nl 
$\Delta C^{c}$ & $-26.1$  & $-10.6$ \nl 
C (d.o.f.) $^{d}$& 518.2(508)& $\ldots$ \nl
\enddata

\tablecomments{\scriptsize Model parameters for the best-fitting
double-Gaussian model for
the double-peaked Fe~K complex in the HEG spectrum IC~4329A
(see \S \ref{doublegauss}). Errors are 90\% 
confidence for one interesting parameter ($\Delta C = 2.706$). 
All parameters are in the source rest frame ($z=0.016054$).
$^{a}$ We show {\it redshift} of the line centroid energy as a
velocity offset when the low and high energy peaks are identified 
with Fe{\sc~i}~K$\alpha$ emission and \felya respectively. 
Fe{\sc~i}~K$\alpha$ emission and \felya emission have rest energies 
of 6.400keV and 6.966keV respectively. Velocities are rounded to the nearest 
5 km $\rm{s}^{-1}$.  Compare these 
values with the systematic uncertainty in the \hetg energy scale of 
$\pm$ 420, $\pm$ 460 km $\rm{s}^{-1}$ at the observed energies
of the emission peaks.
$^{b}$ Gaussian emission-line intensity in units of
$10^{-5}$ photons $\rm{cm}^{-2} \rm{s}^{-1}$. $^{c}$ Change in the
$C$--statistic upon addition of the particular Gaussian model component to a simple 
power-law model for the continuum (there are three additional degrees of freedom for
each Gaussian component).
$^{d}$ Degrees of freedom.} 
\end{deluxetable}

\subsection{Single Relativistic Disk Line Model}
\label{singlediskline}

Next, we replaced the double-Gaussian emission-line model 
described above with a single, broad 
emission line model from a relativistic disk around a Schwarzschild black 
hole (e.g. Fabian \etal 1989).  
These data
cannot distinguish between Kerr and Schwarzschild metrics
(see \S\ref{complexcont}).
The inner radius of the disk was fixed 
at 6$r_{g}$ since it could not be constrained. 
The outer disk radius and line intensity were free 
parameters. The radial line emissivity per unit area was a power law,
$r^{q}$, where $q$ is the emissivity index. 
This time we also included another 
disk-line component to model Fe K$\beta$ emission. Although
it makes little difference to the results, we included
it for completeness. The rest energy was fixed at 7.056 keV (corresponding
to Fe~{\sc i} emission) 
for simplicity even though the energy of Fe $K\alpha$ was allowed
to vary (but the proximity of the latter energy to the Fe~{\sc i}
$K\alpha$ energy, combined with the relative weakness of K$\beta$
easily justifies this assumption).
The Fe{\sc~i}~K$\beta$ disk-line parameters were coupled to those of the 
lower energy $K\alpha$ line, with the intensity 
of K$\beta$ tied to (17/150) times the $K\alpha$ line intensity.
There were a total of five free parameters
for this emission-line model, including the intrinsic rest energy
of the monochromatic emission line. The best-fitting results for this model are given in 
Table~2 and the data and model are shown in
Fig.~4(b). The complex Fe-K emission can clearly be fitted with a single 
broad line given: a flat radial emissivity ($q>-0.72$) with emission 
from $\sim$6--70$~r_{g}$ and $\theta 
\sim 24^{+9}_{-1}$ degrees (90\%, one-parameter uncertainties). 
The best-fitting rest-frame energy for this relativistically
broadened line is 
$6.74^{+0.22}_{-0.13}$ keV, which is 
indicative of He-like Fe.
The $C$--statistic for the single disk-line model
is {\it higher} (by $\sim 8.5$, corresponding to $>99\%$ confidence) 
than that for the double-Gaussian fit in Table~1.
However, the single disk-line model we use 
may be too simplistic (in particular with respect to the
assumed simple emissivity law and the assumption of axisymmetric emission), 
so we cannot rule it out on the basis of the 
fit statistic alone. Note also that the flat line-emissivity 
index may be a result of fixing the inner disk radius at $6r_{g}$.
If the line emission is truncated at a larger radius, a steeper
emissivity law is allowed (e.g. see Done \etal 2000).
When we fixed $q$ at $-2.5$ (a value typically
found in relativistic Fe K line fits to CCD data) and allowed
the inner radius to vary, a fit  was obtained
which was only marginally worse ($|\Delta C|<3$),
and yielded an inner radius of $28^{+11}_{-10} r_{g}$.

\begin{deluxetable}{lrrr}
\tablecaption{Single Disk Line Model and He-like Fe Absorption}
\tablecolumns{4}
%\tablewidth{0pt}
\tablehead{
\colhead{ } & \colhead{Disk line} & \colhead{Disk Line} & \colhead{Absorption} \nl
\colhead{ } & \colhead{(No \fexxvp)$^{a}$} & \colhead{(With \fexxvp)$^{b}$} &
\colhead{Line}  \nl
}
\startdata
E(keV)& $6.74^{+0.22}_{-0.13} \ ^{c}$ & $6.85_{-0.17}^{+0.08}$ & 6.7 (fixed) \nl 
$q$ &$>-0.7$ & $0.6^{+3.3}_{-1.7}$ & \dots  \nl 
$\theta
(^{\circ})$& $24^{+9}_{-1}$ & $22^{+7}_{-12}$ & \ldots \nl 
$r_{\rm out}/r_{g}$ &
$55^{+9}_{-9}$ & $48^{+24}_{-25}$ &  \nl 
FWHM($\rm km \ s^{-1}$) & \ldots & \ldots & $12,120^{+11,070}_{-6530}$ \nl
I $^{d}$ & $23.7^{+7.0}_{-6.1}$ & $33.8^{+7.4}_{-11.1}$ & \ldots  \nl
 EW(eV) &
$161^{+48}_{-41}$ & $249^{+55}_{-82}$ & $46^{+22}_{-28}$ \nl 
$\Gamma$& $1.74^{+0.06}_{-0.05}$ & $1.75^{+0.06}_{-0.05}$ & \ldots \nl 
$\Delta C ^{e}$ & 32.3 & \ldots & \ldots \nl 
$C$ (d.o.f.) $^{f}$& 526.7 (509) & 520.5 (507) & \ldots \nl
\enddata
\tablecomments{Model parameters for the best-fitting 
relativistic Schwarzschild disk line models 
(with and without \fexxv He-like resonance absorption) fitted to the Fe~K complex 
in the HEG spectrum of IC~4329A (see \S \ref{singlediskline}).
All parameters are in the source rest frame ($z=0.016054$).
Errors are 90\% confidence for one 
interesting parameter ($\Delta C = 2.706$). The inner radius of the 
disk-line model component was fixed at 6$r_{g}$. 
$^{a}$ Model fitted with {\it no} \fexxv He-like resonance absorption.
$^{b}$ Model includes an inverted Gaussian with center energy
fixed at 6.7~keV to model \fexxv He-like resonance absorption
(best-fitting absorption-line parameters are given in the last column
of the table).
$^{c}$ There are local minima all the way up to the rest energy of \felya (6.966~keV).
$^{d}$ Fe-K emission-line intensity
in units of $10^{-5}$ photons $\rm{cm}^{-2} \rm{s}^{-1}$. 
$^{e}$ Change in the $C$-statistic upon the 
addition of the disk-line model component to a simple power-law model. 
$^{f}$ Degrees of freedom.}
\end{deluxetable}

\subsection{Single Disk Line with \fexxv He-like Absorption}
\label{absorptionline}

It is conceivable that between the low-energy and high-energy
peaks of the Fe K emission-line complex, we may have \fexxv He-like
resonance absorption at 6.7~keV. Such resonance absorption cutting away
the Fe K emission-line profile has been suggested for NGC~3516 (Nandra 
\etal 1999) and MCG~$-$6-30-15 (Fabian \etal 2002) and discussed
in a theoretical context by, for example, Ruszkowski \& Fabian (2002).
Therefore, we tested the data against a simple model of
\fexxv He-like resonance absorption by adding an inverted Gaussian
component to the single disk-line model described in \S\ref{singlediskline}.
The center energy was fixed at 6.7~keV in the rest frame of the
source, which meant that two more free parameters were added to
the model (the width and equivalent width of the absorption line).
The results for both the relativistic Fe K emission line and the
absorption line are shown in Table~2 so that they can be
directly compared with the fit with the relativistic Fe K emission line
only (\S\ref{singlediskline}). It can be seen that the addition of
the absorption line has little effect on the disk inclination angle,
the disk outer radius, and the radial emissivity derived from
the broad Fe K line. Naturally, the intensity and equivalent width
of the Fe K emission line are required to be larger due to the
presence of the absorption line. The derived equivalent width
of the absorption line is $46^{+22}_{-28}$~eV, and its width
is $12,120^{+11,070}_{-6530} \rm \ km \ s^{-1}$. These values
depend somewhat on how the underlying emission line is modeled.
However, since the addition of the absorption line decreases
the $C$--statistic by only 6.2 for the addition of two
free parameters (corresponding to a significance of $<99\%$),
we do not pursue the matter further.

\subsection{Double Disk Line Model}
\label{doubledisk}

Next, we fit the Fe~K emission-line 
complex with two $K\alpha$ disk-line model components.
The idea here is to investigate
in a very simplistic way, the
possibility of line emission from a disk with a vertically stratified
ionization structure. For example, the higher energy peak
could come from a highly ionized top layer, whilst the
lower energy line emission could come from a cooler layer
beneath the hot layer. Pairs of parameters
corresponding to the inclination angle, the inner
and outer radii for the two
disk-line components were tied so that they were
forced to vary together.
Both disk-line model components had inner radii fixed at 6$r_{g}$.
The rest-energy for the lower energy line component was
a free parameter but the rest energy of the higher energy line was
fixed at the rest--energy
of \felya (6.966 keV).
Thus, the radial line emissivities of the
two disk-line components were independent. 
The above simplifying assumptions are necessary, given the
limitations of the data, in order not to over-parameterize the
model.
Again, for completeness we added an extra disk-line component 
to model \fekb emission associated with the lower energy
peak, as described in \S \ref{singlediskline} for the
single disk-line model. However, inclusion of the \fekb emission
does {\it not} add any more free parameters to the model.
Thus, there were a total of six free 
parameters in this model of the Fe K emission complex.
The 
best-fitting results for this model are given in Table~3
and the data and model are shown in Fig~4(c). Clearly 
both disk lines have very similar radial emissivity 
profiles and are strongly constrained to a near face-on inclination 
($\theta < 6.0^{\circ}$ at $>95\%$ confidence). The EWs of the 
low and high energy features, as modeled by disk lines, are larger 
than the corresponding EWs in Table~1 (double-Gaussian model) and their combined 
EW is larger than the single disk-line fit to the Fe~K region (at 90$\%$ 
confidence). The fit statistic is an improvement over the double-Gaussian 
model fit at $>95\%$ confidence and is an improvement over the single 
disk-line model at $>3 \sigma$ confidence ($\Delta C =-12.6$ for one 
additional free parameter). 

\begin{deluxetable}{lrr}
\tablecaption{Double Disk-Line Model}
\tablecolumns{3}
\tablewidth{0pt}
\tablehead{
\colhead{ } & \colhead{Low-Ionization Fe K } & 
\colhead{Fe~{\sc xxvi} \ Ly$\alpha$}\nl}
\startdata
E(keV) & $6.443^{+0.050}_{-0.016}$ & $6.966$ fixed \nl
$q^{a}$ &$-2.44^{+0.22}_{-0.24}$ & $-2.20^{+0.33}_{-0.37}$ \nl 
$\theta(^{\circ})$$^{b}$& $<6.0$ & $\ldots$ \nl 
$r_{\rm out}/r_{g} ^{b}$ &$633^{+57}_{-369} $ & $\ldots$ \nl 
$I \ ^{c}$ & $20.6_{-6.2}^{+6.4}$ 
& $14.5_{-7.0}^{+6.3}$ \nl 
EW(eV)& $120^{+37}_{-36}$ & $101^{+44}_{-49}$\nl 
$\Gamma$& $1.79_{-0.05}^{+0.06}$ & $\ldots$ \nl 
$\Delta$ C$^{d}$ & $-25.2$ & $-22.9$ \nl 
$C$ (d.o.f.) $^{e}$& 514.1 (507) & $\ldots$ \nl
\enddata
\tablecomments{\scriptsize Parameters for the best-fitting double 
relativistic disk-line fit to the Fe~K line region of the HEG spectrum 
in IC~4329A. 
The model includes an
Fe~K~$\beta$ emission component for the lower-energy line (see \S \ref{doubledisk}
for fitting details).
Errors 
are 90\% confidence for one interesting parameter ($\Delta C = 2.706$)
except where noted. 
The inner radius of all disk-line model components was fixed at 6$r_{g}$. 
All parameters are in the source rest frame ($z=0.016054$).
The energy of the Fe~{\sc xxvi}~Ly$\alpha$ 
disk line component was fixed at its rest energy of 6.966~keV. $^{a}$ 
Power-law index for the line radial emissivity. The lower energy Fe~K$\alpha$ and 
Fe~{\sc xxvi} Ly$\alpha$ disk line components were coupled in all 
parameters except this one ($q$).
$^{b}$ Both $\theta$ and $r_{\rm out}/r_{g}$ for the
\felya disk line were forced to their corresponding values for
the lower energy disk line during the fitting.
The $C$-statistic varies non-monotonically close to the best fit.
For the disk inclination angle the $95\%$ 
confidence upper limit is shown. Formally, $\theta<0.35^{\circ}$ at $90\%$ confidence but 
the $95\%$ confidence value is more reliable since it is in the
region where $C$ is well-behaved. For $r_{\rm out}/r_{g}$, the
errors represent $90\%$ confidence values, but only when $\Delta C$ never
drops below 2.706 again.
$^{c}$ Line intensity in units of $10^{-5}$ photons $\rm{cm}^{-2} \rm{s}^{-1}$. 
$^{d}$ Change in C--statistic upon the addition of the 
particular disk line 
model component to a power-law continuum only.
$^{e}$ Degrees of freedom.} 
\end{deluxetable}

\subsection{Relativistic Disk Line and Gaussian Model}
\label{diskgauss}

Next, we modeled the two peaks in the Fe~K band with a 
relativistic disk line for the lower energy
emission component and a Gaussian for the higher energy component.
Again, we included Fe~K$\beta$ emission as described in \S \ref{singlediskline}
for completeness.
The best-fitting results for this model are given in Table~4.  
The disk inclination angle is tightly constrained to be 
$12 \pm 2$ degrees and $q$ is flat. 
The Gaussian best-fitting parameters  to 
the high-energy feature are similar to those in Table~1, as we should 
expect. This model is not significantly better than the double-Gaussian 
model statistically (Table~1), but it is better than the single broad 
disk-line fit (Table~2) at $>95\%$ confidence since the C--statistic 
decreases by $\sim 10.3$ for three additional free parameters
(relative to the single disk-line model). 
Note that the radial emissivity power-law index ($q$) for the
disk-line emission is very flat.

\subsection{Complex Continuum}
\label{complexcont}

So far, we have not included the effects of a Compton-reflection
continuum, or indeed any other complexity in the continuum.
First of all, we note that there is little evidence of an
Fe~K edge above 7~keV in the data (Fig.~4). An Fe~K edge could be
produced either in transmission or reflection from optically-thick
matter. We added a simple edge model to the disk-line plus Gaussian
model (\S \ref{diskgauss}), at 7.11~keV (corresponding to neutral
Fe) and found a best-fit value of zero, and a 90\%, one-parameter
upper limit of 
$\tau_{Fe}<0.054$. This is
comparable with $\tau_{Fe} \sim 0.03$ from the \xmm observation of 
\ic by Gondoin \etal (2001). This implies a total neutral absorbing 
column of $N_{H}<4.9 \times 10^{22} \ \rm{cm}^{-2}$ in the 
line-of-sight. This is certainly consistent with the low-energy
MEG data (McKernan, Yaqoob, \& Reynolds 2004, in preparation).

Due to the restricted bandpass and limited signal-to-noise, the
data are not sensitive to the Compton-reflection continuum, which
becomes important above $\sim 7$~keV. However, we demonstrated
this directly by fitting a model consisting of a
power law with a high-energy exponential cut-off, and
Compton reflection from optically-thick,
neutral matter (using the XSPEC model {\tt pexrav}, as
described in Magdziarz \& Zdziarski 1995), a \fekalfa line
to model the lower energy peak in the Fe complex, and another
emission line to model the higher energy peak. A Fe~$K\beta$ line
was also included, for completeness, and tied to the
\fekalfa line as described in \S \ref{singlediskline}.  
The Fe~$K\alpha$ line, the Fe~$K\beta$ line, and the Compton reflection
continuum were all blurred with a relativistic kernel based on
the Kerr metric (Laor 1991; see also Fabian \etal 2002). 
The reason for now using the Kerr metric (for a maximally spinning
black hole), instead of the Schwarzschild metric, is to directly
demonstrate that the data are not sensitive to the black-hole
angular momentum. Since
the higher energy peak of the Fe complex (possibly due to \felyap) 
in the IC~4329A HEG data
is quite narrow, it was modeled with a Gaussian and {\it not}
subjected to the relativistic blurring.
Also, we included no Compton-reflection continuum
associated with the the higher energy peak. This is because
if it is due to highly ionized Fe, the curvature 
in the reflected continuum is much less than the case for
a disk which is not ionized, especially for data in the
restricted bandpass (3--8~keV) that we are dealing with here. 
If the higher energy feature is not
a separate line, then the single Compton-reflection continuum
already included is adequate.

The free parameters of the {\tt laor} model were the disk inclination angle,
$\theta$,
the power-law emissivity index, $q$, and the outer disk radius, $r_{\rm out}$.
Since even the lower energy emission peak is not extremely
broad, we found the fit to be quite insensitive to the
inner disk radius, so again fixed it at $6r_{g}$.
The Compton-reflection model ({\tt pexrav}) does not allow the
disk inclination angle to be tied to that of the {\tt laor} model,
and also does not allow inclination angles $<18^{\circ}$. Therefore
we fixed the inclination angle at $30^{\circ}$ for the Compton-reflection
continuum. This is adequate for our purpose. We also assumed 
solar element abundances and an exponential cut-off energy of 200~keV
for the intrinsic power-law continuum. This is consistent with
broadband observations of IC~4329A as well as
results for other AGNs (e.g. Madejski
\etal 1995; Malizia \etal 2003; Perola \etal 2002). In any case, 
our conclusions are not sensitive to this assumption. 
The Compton-reflection continuum then had only one free parameter in the fit,
the ratio, $R$, of the amplitude of the
reflected continuum to that expected from a centrally illuminated
infinite disk, subtending a solid angle of $2\pi$ at the X-ray source. 
In all, this model had eleven free parameters, including the overall
continuum normalization. This is one more free parameter than
the equivalent model that utilized the
Schwarzschild metric and that did not include
a Compton-reflection continuum (i.e. the model described in \S 
\ref{diskgauss}).

\begin{deluxetable}{lrrrr}
\tablecaption{Disk Line Plus Gaussian Line Models}
\tablecolumns{5}
%\tablewidth{0pt}
\tablehead{
\colhead{ } & \colhead{Disk Line} & \colhead{Gaussian} &
\colhead{Disk Line} & \colhead{Gaussian} \nl
%\colhead{ } & \colhead{feature} & \colhead{feature} & \colhead{feature}\nl}
\colhead{ } & \colhead{(simple)$^{a}$} & \colhead{(simple)$^{a}$} &
\colhead{(complex)$^{b}$} & \colhead{(complex)$^{b}$}  \nl
}
\startdata
E(keV)& $6.412^{+0.045}_{-0.022}$ & $6.900^{+0.036}_{-0.030}$ &
$6.349^{+0.262}_{-0.071}$ &
$6.900^{+0.036}_{-0.030}$\nl
$q$ &$-0.37^{+2.50}_{-1.08}$ & \ldots & $>-2.4$ & $\ldots$ \nl 
$\theta (^{\circ})$& $12 \pm 2$ & \ldots & $18^{+7}_{16}$ & $\ldots$ \nl 
$r_{\rm out}/r_{g}$ & $105^{+31}_{-79}$ & \ldots & $>13$ &
$\ldots$ \nl 
c$\Delta$E/$E_{0}$(km $\rm{s}^{-1}$)$^{c}$ &$\ldots$ & 
$2840^{+1295}_{-1550}$ & \ldots & $2840^{+1295}_{-1550}$ \nl 
$\sigma$(eV) & $\ldots$ & 
$46^{+36}_{-31}$ & \ldots & $40_{-27}^{+41}$ \nl 
FWHM(km $\rm{s}^{-1}$) $^{c}$ &$\ldots$ & 
$4710^{+3680}_{-3175}$ & \ldots & $4095^{+4195}_{-2765}$ \nl 
I$^{d}$
 & $16.1^{+5.0}_{-4.8}$ & $5.4^{+5.1}_{-3.2}$ & 
$15.9^{+13.7}_{-6.9}$& $6.0_{-2.9}^{+3.3}$ \nl 
EW (eV)&
$96^{+30}_{-29}$ & $36^{+34}_{-21}$ & $88^{+76}_{-38}$ & $40^{+22}_{-21}$ \nl 
$\Gamma$& $1.75 \pm 0.05$ & \ldots & 
$1.87^{+0.20}_{-0.17}$ & $\ldots$ \nl 
$R^{e}$ & \ldots & \ldots & $<2.9$ \ldots \nl
$\Delta C$ & $-30.2^{g}$ & $-12.2^{h}$ & $-32.6^{g}$ & $-9.6^{h}$ \nl 
$C$ (d.o.f.)$^{f}$& 516.4 (506) & \ldots & 516.8(505) & $\ldots$ \nl
\enddata
\tablecomments{\scriptsize Model parameters for the best-fitting
relativistic disk line plus Gaussian models to the Fe~K complex in the HEG
spectrum of IC~4329A (see \S \ref{diskgauss} for fitting details). 
The model also includes Fe~K~$\beta$ emission
for the lower energy line (see \S \ref{singlediskline} for details).
Errors are 90\% 
confidence for one interesting parameter ($\Delta C = 2.706$). The inner 
radius of the lower energy Fe~K$\alpha$ disk-line model was fixed 
at 6$r_{g}$. All quantities are quoted in the source rest frame
($z=0.016054$). $^{a}$ Simple model with a power-law continuum, and a
relativistic disk line for a Schwarszchild black-hole. 
$^{b}$ Relativistic disk line from maximally rotating Kerr black-hole,
with a power-law plus Compton-reflection continuum blurred by the 
relativistic line kernel. 
Fe~{\sc xxvi} Ly$\alpha$ has a rest energy of 6.966~keV. 
$^{c}$ Velocities are rounded to 
nearest 5 km $\rm{s}^{-1}$. 
$^{d}$ Line intensity in units of $10^{-5}$ photons $\rm{cm}^{-2}
\rm{s}^{-1}$. $^{e}$ 
Effective `reflection fraction' (i.e. $R=1$ corresponds
to the steady-state Compton reflection amplitude from an infinite
disk subtending $2\pi$ solid angle at the X-ray source). $^{f}$
Degrees of freedom. $^{g}$
Change in the $C$--statistic upon addition of only the 
disk line to the continuum. $^{h}$ 
Change in the $C$--statistic upon addition of a Gaussian 
emission-line component
to the disk-line plus continuum model.}
\end{deluxetable}

The results are shown in Table~4 so that they can be
directly compared with the results when Compton reflection was
not included, and the continuum was modeled only with a simple power law.
It can be seen that the quality of the fits for the two models,
the best-fitting values for the emission-line parameters
and their statistical error ranges are virtually indistinguishable.
This applies to both the relativistic and non-relativistic emission lines.
It is important to note, however, that inclusion of the Compton-reflection
continuum does allow the line radial emissivity index to be steeper,
relieving the requirement that the emissivity law be flat (which
may be physically difficult to achieve).
Another noticeable difference between the simple
and complex continuum fits is that the latter fit yields a somewhat steeper
intrinsic X-ray continuum power-law index.
For the fit with a complex continuum, the
reflection fraction, $R$, was consistent with zero, and we obtained
an upper limit of $R<2.9$ (90\% confidence, one parameter). This 
confirms  that the data are insensitive to the Compton-reflection continuum.
The result also justifies using only a simple power-law for the 
continuum in the model-fitting described in \S \ref{doublegauss} --
\S \ref{diskgauss}.
Since the relativistically smeared Compton-reflection continuum
is smoother and produces less curvature in the net spectrum
than an unsmeared continuum
(in the energy band relevant here), we tested the data against 
the same model with the smearing of the Compton-reflection component
turned off. Physically, such a component could originate in 
optically-thick distant matter, such as a parsec-scale torus.
Again, no reflection
was formally required by the data, with a 90\% confidence upper limit of 
$R<1.7$. These results further justify using only a simple power law
for the continuum in the model-fitting described in \S \ref{doublegauss} --
\S \ref{diskgauss}.

\section{Discussion}
\label{sec:discussion}
In a \chandra \hetg observation of IC~4329A, we have measured
broad and complex line emission in the Fe K band
using the
highest spectral resolution currently available in the Fe K band ($\sim 1860 \ \rm km
\ s^{-1}$ FWHM at 6.4~keV).
The line profile is double-peaked, the peaks occurring at $\sim 6.3$~keV
and $\sim 6.9$~keV in the source rest frame.
Modeled with Gaussians, the lower and higher energy peaks have widths of 
$20,825^{+10,110}_{-7,375}$ km $\rm{s}^{-1}$ FWHM 
and $3680^{+5200}_{-2455}$ km $\rm{s}^{-1}$ FWHM 
respectively. The lower energy peak is resolved by the
high energy grating (HEG) at $>99\%$ confidence, whilst the higher energy peak is
only resolved at $<90\%$ confidence. 
The low-energy peak is redshifted from the expected rest-frame
energy of Fe~{\sc~i}~K$\alpha$ by at least $650$ km $\rm{s}^{-1}$ and
the high-energy peak is redshifted from the expected rest-frame
energy of \felya by at least $950$ km $\rm{s}^{-1}$. Both of these
lower limits on the redshift correspond to 90\% confidence levels for
one parameter. 
The emission-line peak at $\sim 6.9$~keV that we found in the
\chandra \hetg data has not been detected
in any previous observations of IC~4329A.
We also detected a much weaker emission peak at the
expected energy of the \fexxv forbidden line energy at the
systemic velocity of IC~4329A, but the
feature is detected only with a marginal significance ($>95\%$ confidence).

The amount of Fe in the line-of-sight may be insufficient to produce all 
of the 
observed EW of the lower energy $\sim 6.3$~keV 
Fe~K feature ($110^{+46}_{-40}$ eV from 
Table~1) since the measured (HEG)
upper limit on the optical depth of an Fe~K edge
corresponds to an absorbing column density $<4.9 \times 10^{22} \rm{cm}^{-2}$ 
and this is consistent with the low-energy MEG
data from the same observation (McKernan, Yaqoob \& Reynolds 2004 in 
preparation; see also Gondoin \etal 2001). Such a column density
can at most produce an EW of $<50$~eV, even if 
it fully covered the sky (e.g. Krolik \& Kallman, 1987; 
Yaqoob \etal 2001). To investigate the contribution to the Fe~K line emission
from distant matter, we added an extra (narrow) Gaussian line component to
the two Gaussian model described in Table~1. The energy of the new Gaussian
component was fixed at 6.4 keV in the rest--frame of the AGN since we are
testing the data for a line which originates from matter sufficiently distant
that there is no gravitational redshift. Also, a line originating from such
distant matter would be unresolved by \chandra, so we fixed the width of the
Gaussian at a value (1 eV) much less than the instrument resolution. The two
other Gaussian model components were allowed to vary as before. The 
best--fitting model did not result in a significant improvement in the 
fit--statistic ($\Delta C=2.3$ for one additional free parameter).
 Thus, the values and error ranges of all the model parameters did not change
significantly compared to the values shown in Table~1, which were obtained 
without the distant--matter line. The $90\%$ confidence (one--parameter) 
upper limit on the distant--matter line EW we obtained is 18 eV, so that 
it could contribute at most $\sim 1/4$ of the low energy (6.2--6.4 keV) Fe~K
 line emission in IC~4329A. This and the fact that the low-energy peak 
is at 6.3~keV and resolved by the HEG
suggests that the contribution from distant matter is not
as important as in some other AGN in which the Fe K line
emission peaks at 6.4~keV and is unresolved by the HEG (see Yaqoob
\& Padmanabhan 2004). Thus, it is likely that the bulk of the $\sim 6.3$~keV 
line emission originates in an accretion disk. It is difficult
to deduce what the ionization state of the matter must be
because of the possibility of a peak at energies higher than
6.4~keV being gravitationally redshifted down to $\sim 6.3$~keV.
Note that a radius of $\sim 65r_{g}$ is required to gravitationally
redshift a 6.4~keV line to 6.3~keV.

The interpretation of the overall, double-peaked Fe K line
profile is difficult. Several physically distinct models can describe
the data. The two peaks may correspond to the
Doppler horns of a single disk line or they may be artificial in the
sense that He-like Fe K absorption has removed the central
portion of a disk line profile which may not have prominent Doppler peaks.
Alternatively, the higher energy peak at $6.9$~keV may correspond
to \felya emission which is independent of the line emission associated
with the lower energy peak. In that case, the \felya line may or may not
originate in a disk. Again the two scenarios are
degenerate as far as the data are concerned.
If there is \felya  line emission from a disk it may come from
a layer which is hotter than that producing the lower-energy line
(for example the hot corona, e.g. see Nayakshin \&
Kallman 2001; Ballantyne, Ross, \& Fabian 2001). 
Although the $\sim 6.9$~keV line appears to be 
fairly narrow we have shown by direct
model fitting that it could still originate from 
a disk which extends down to $6r_{g}$, with a radial line emissivity per unit area
which is {\it not} flat.

The line emission peak at $6.9$~keV is reminiscent of a similar
feature found in a long \xmm exposure of the Seyfert galaxy
MCG~$-$6-30-15 (e.g. Fabian \etal 2002). In that case there
is also an ambiguity between \fexxv He-like absorption and 
and \felya emission, and the peak energy is slightly redshifted
from the expected energy of \felyap, just as in IC~4329A.
\fexxv He-like absorption has actually been observed with
less ambiguity in some AGN (PG~1211$+$143, Pounds \etal 2003a;
PG~0844$+$349, Pounds \etal 2003b; NGC~3783, Reeves \etal 2003).

Although there are several interpretations of the complex
Fe K emission revealed by the HEG spectrum of IC~4329A,
each of the different scenarios is interesting from the point of
view that future observations with {\it Astro-E2}
will break much of the degeneracy and
whichever scenario turns out to be correct,
we would learn 
something new about the central engine. In particular, 
observational constraints on the ionization structure of
the accretion disk in AGN are currently very poor and
IC~4329A may be a good astrophysical laboratory for improving them.

We gratefully acknowledge support from NSF grant AST0205990 (BM) and 
NASA grants NCC5-447 (TY) and NAG5-10769 (TY).  
We thank A. C.~Fabian for providing us with
a relativistic blurring routine.
We also
thank Urmila Padmanabhan for help with some of the data analysis.
We made use of the 
HEASARC on-line data archive services, supported by NASA/GSFC and also 
of the NASA/IPAC Extragalactic Database (NED) which is operated by the 
Jet Propulsion Laboratory, California Institute of Technology, under 
contract with NASA.  We are grateful to the \chandra instrument and 
operations teams for making this observation possible.

\end{document}